# Synthesis of graphene aerogels toward the absorption of liquid and gas pollutants

G. Gorgolis[1,2], N. Pastra[1], G. Paterakis[1,2] and C. Galiotis[1,2,*]

[1] Faculty of Chemical Engineering – University of Patras, Panepistimioupoli – 26504 Rio – Greece
[2] Foundation of Research and Technology – Institute of Chemical Engineering Sciences, Stadiou Str. – Platani – 26504, Greece

* galiotis@chemeng.upatras.gr

## Abstract

Environmental disasters, such as the oil spill in Mexico gulf or, more recently, the Mauritius oil spill caused by a bulk carrier vessel, are some of the ecological issues that the modern society faces frequently. These accidents highlight the need for development of efficient materials that can be employed to eliminate pollutants including crude oil and its derivatives, as well as toxic organic solvents. The objective of the present study is to achieve the synthesis of graphene aerogels with two main methods and to determine their efficiency as absorbents of liquid and gas pollutants. Graphene aerogels were prepared according to the information given by Hong et al. [1] and Yang et al. [2]. Both methods are based on the chemical reduction of graphene oxide as the first step and the formation of the aerogel as the second, providing a more cost-effective solution and resulting in a lattice structure similar to graphene. The fabricated materials were investigated for their absorption capacity for liquid and gas pollutants. The specimens were mounted in airtight microenvironments saturated with a specific Volatile Organic Compound (VOC). Three VOCs were tested; formaldehyde, acetic acid and hydrochloric acid and the results were based on gravimetric measurements. In order to increase the accuracy of the obtained data, a VOCs sensor was used to measure the change of voltage before and after the introduction of graphene aerogels in a saturated microenvironment. Furthermore, the graphene aerogels were evaluated as absorbents in organic solvents, namely acetone, ethanol and methanol.

## 1. Introduction

Graphene is a carbon-based nanomaterial commonly referred as a two-dimensional $sp^2$ bonded sheet-like material and it serves as a structural element of carbon allotropes, such as graphite, charcoal, carbon nanotubes and fullerenes. As a 2D material it has thickness equivalent to an atom diameter, while on the other two dimensions it stretches indefinitely. Each carbon atom forms σ bonds with three neighboring atoms and as a result graphene is arranged in a hexagonal lattice (honeycomb). The remaining electron of the carbon atom forms a $\pi$ orbital that contributes to a delocalized network of electrons [3]. Due to the unpaired p-orbitals graphene can be characterized as a semimetal. The lateral dimension of the microscopic corrugations is estimated at 8-10 nm, while the height displacement is estimated at 0.7-1 nm (Fig. 1.1). There is a symmetry between sublattices A and B, which lead to the degeneration of the conduction and valence states at the K and K' points of the Brillouin zone and the subsequent rise to the linear dispersion of electronic bands (Fig. 1.1.c). Near the two inequivalent graphene valleys the electrons behave as massless Dirac fermions with velocity $v'' \cong 8\ cm\ s)*$ [3].

Considering the synthesis of graphene and its derivatives, the desired structure and properties are largely dependent upon the size, shape and functional groups attached to the surface of the material. The bottom-up process has been proven difficult for industrial applications, while the top- down approach is easier to generate GO, with both $sp^2$ and $sp^3$ carbons containing abundant oxygen groups, which upon reduction (rGO) can eliminate most of the oxygen groups and sp3 carbon to generate more graphene-like material with much improved properties. GO synthesis can be divided into two main categories: bottom up methods where simple carbon molecules are used to construct pristine graphene, and top-down methods where layers of graphene derivatives are extracted from a carbon source, typically graphite [4]. Aside from the operative oxidative mechanisms, the precise chemical structure of GO has been debated over the years, and even to this day there are only ambiguous models. The primary obstacles on the modelling process are the complexity of the material, the lack of consistency from sample-to- sample, because of its amorphous character and the nonstoichiometric atomic composition. The most well-known model is the one by Lerf and Klinowski [3]. According to this model, the carbon plane is laced with hydroxyl and epoxy (1,2 ether) functional groups, while carbonyl groups are present most likely as carboxylic acids along the sheet. The GO sheets consist partly of tetrahedrally bonded sp3 carbon atoms, which are displaced slightly above or below the graphene plane. Due to the structure deformation and the presence of covalently-boned functional groups, GO sheets have shown an anatomical roughness [4].

In order to produce materials with properties as close to graphene as possible, research has been carried out to reduce the oxygen functional groups of GO. This reduction is accomplished by a number of processes, from thermal to electrochemical, each of which leads to differences in the proper- ties [5]. The key design factors in GO reduction include the C/O ratio of the end product, selectivity in every single type of oxygen group, healing of the surface defects of the GO from oxidation and choice of green reducing agents, as well as maintaining the desired physical and chemical properties of the GO, such as strength, conductivity, optical properties and solubility/dis- solubility [3].

GO is the appropriate precursor to prepare 3D graphene assemblies due to its high dispersion in aqueous media and its functionality. Also, it is very useful in synthesizing graphene aerogels, because the oxygen from the functional groups of both basal planes and edges are able to react covalently with different compounds and thus, yield new materials with properties that can be tailored to specific applications. The 3D assembly of GO is achieved by introducing chemical agents whose repulsive forces, in combination with applied heat lead to the formation of aerogels. The synthesis of 3D graphene-based materials derives from the need to exploit the properties of graphene at a larger scale and via facile preparation methods. Graphene aerogels (GAs) have been investigated for their significantly low density, high electrical conductivity, organic dye pollutant removal and other applications. According to literature [6], [7] there are various categories of synthesis approaches with the most principal being the oxidation- reduction (chemical, hydrothermal, electrochemical or template-directed) and cross-linking methods.

The term freeze drying refers to the process of water removal from a material and entails the freezing of the material followed by pressure reduction combined with the supply of heat, in order to allow the sublimation of the frozen water [8]. It is typically used for the preservation of perishable materials, the extension of shelf-life or to facilitate their transportation in the food and pharmaceutical industries. The end products of freeze-drying are stable in ambient temperature, consequently counterbalancing the high cost of machinery and the lack of need for refrigeration.

## 2. Experimental procedure

### 2.1 Graphene Oxide preparation

The graphene oxide (GO) is synthesised from natural graphite flakes (NGS Naturgraphit GmbH, Germany) by a two-step oxidation process, based on a method which is a modification of Hummer's method, namely, a pre-oxidation step at the beginning, followed by the final oxidation step where GO flakes are collected in an aqueous dispersion (analytical information can be found in the SI file).

### 2.2 Freeze-drying technique

In the experimental setup of the graphene aerogel freeze drying the freeze dryer used can follow both the manifold and batch method. In the thesis the batch method was used. That means that a big number of vessels with similar size are placed together in a tray dryer. Almost all products are processed in the same conditions of temperature and pressure, which can be controlled throughout the drying with precision. Some differences in heat input may cause variations in the residual moisture of the end product. Specifically, the Telstar Cryodos is a benchtop laboratory freeze-dryer whose main characteristics are the standard chamber of 3 thermostatic shelves and the 8-port manifold with valves for flasks and wide-neck filter bottles (Progen Scientific, 2017). It also has a double stage vacuum pump with gas ballast and an exhaust filter, as well as a safety valve for the system's isolation. The vacuum pump is located outside the cabinet. A refrigerating system, a cascade system, a coil type condenser and two hermetic compressors with cascade mounting are all included. The total condenser capacity is 8 kg, while the heating of the shelves can reach a temperature of 70°C (Telstar, 2020), while the temperature goes down to -85°C.

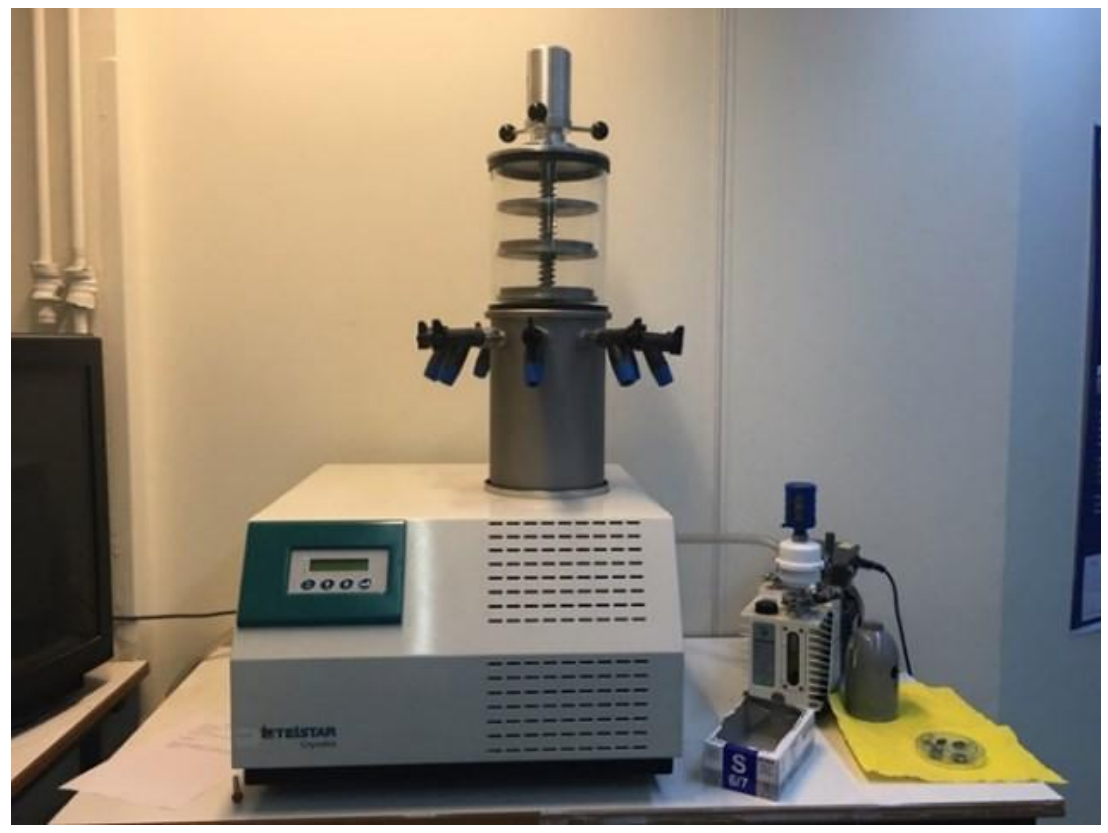

Figure 1: Laboratory freeze dryer used for the preparation of the graphene aerogels with the freeze-drying method.

## 2.3 Graphene Aerogels synthesis

The approach presented previously by Hong et al. [1] was followed herein for the synthesis of the reduced graphene oxide (rGO) aerogels. Aqueous solution of GO after its preparation, was subsequently diluted in water to obtain a concentration of 1 mg/mL. Hypophosphorous acid ($H_3PO_2$) and iodine ($I_2$) of weight ratio GO:$H_3PO_2$:$I_2$ 1:100:10, were then added as the chemically reducing agents. Subsequently, the solution was placed in a furnace and heated to 80°C for 8 hours, resulting in a uniform gelation of the GO. The sample was then rinsed with water until a pH of equal to 7, followed by freeze-drying for 48 h. The chemical reaction that relates to the chemical reduction of graphene oxide has as reactants both $H_3PO_2$ and $I_2$. Based on the proposed self-assembly mechanism of the graphene sheets, at first, $H_3PO_2$ is combined to $I_2$ and creates hydriodic acid that acts as the reducing agent and is responsible for the pH decrease of the medium (pH<1). The decreased value of pH results in diminished electrostatic repulsions, which then aids the formation of GO agglomerates. The oxygen-containing functional groups on the GO surface are reduced from hydriodic acid and produce chemically converted graphene. With the aim to create a more stable structure, the rinsed hydrogel was pre-freezed in a refrigerator (-27°C) up to a week until the freeze-drying took place (**Figure S1**).

## 2.5 Structural Characterisation Methods

SEM photos were taken using a LEO SUPRA 35 VP, while Raman spectra of the specimens were recorded using an InVia Reflex (Renishaw, UK) MicroRaman equipment using a 633 nm laser excitation. In all experiments, spectra were recorded at several points on each specimen using a Renishaw InVia Raman Spectrometer with a 1200 groove/mm$^{-1}$ grating and a x100 lens. The power of the laser beam was kept below 1 mW to avoid heating of the specimen. Raman spectra were baseline corrected and graphene peaks were fitted to Lorentzian functions. When graphene peaks were superimposed onto the peaks of the substrates, the necessary deconvolution process was applied. In this analysis, the Lorentzian components assigned to the substrates were held fixed, having had their parameters (position: full-width at half-maximum) evaluated from the spectra of the bare substrates. X-ray diffraction measurements were performed with the assistance of a Bruker D8 Advance model diffractometer.

The surface analysis measurements were performed in a UHV chamber (P~$5\times10^{-10}$ mbar) equipped with a SPECS Phoibos 100-1D-DLD hemispherical electron analyser and a non-monochromatized dual-anode Mg/Al x-ray source for XPS. The XP Spectra were recorded with MgKa at 1253.6 eV photon energy and an analyser pass energy of 15 eV giving a Full Width at Half Maximum (FWHM) of 0.85 eV for Ag3$d_{5/2}$ line. The analysed area was a spot with the diameter of 3 mm. The atomic ratios were calculated from the intensity (peak area) of the XPS peaks weighted with the corresponding relative sensitivity factors (RSF) and the energy analyser transmission function. For spectra collection and treatment, including fitting, the commercial software SpecsLab Prodigy (by Specs GmbH, Berlin) was used. The XPS peaks were deconvoluted with a sum Gaussian-Lorenzian peaks after a Shirtey type background subtraction.

## 3. Results and Discussion

### 3.1 CHARACTERIZATION OF GRAPHENE OXIDE

The GO 3.2 mg/mL that was used during the entirety of experiments was characterized by UV-VIS spectroscopy. Specifically, aqueous solutions of GO were prepared with different compositions. Each solution was prepared with the same GO quantity as a base for the comparison between the level of dilution and the observed absorbance on the UV-Vis spectrum.

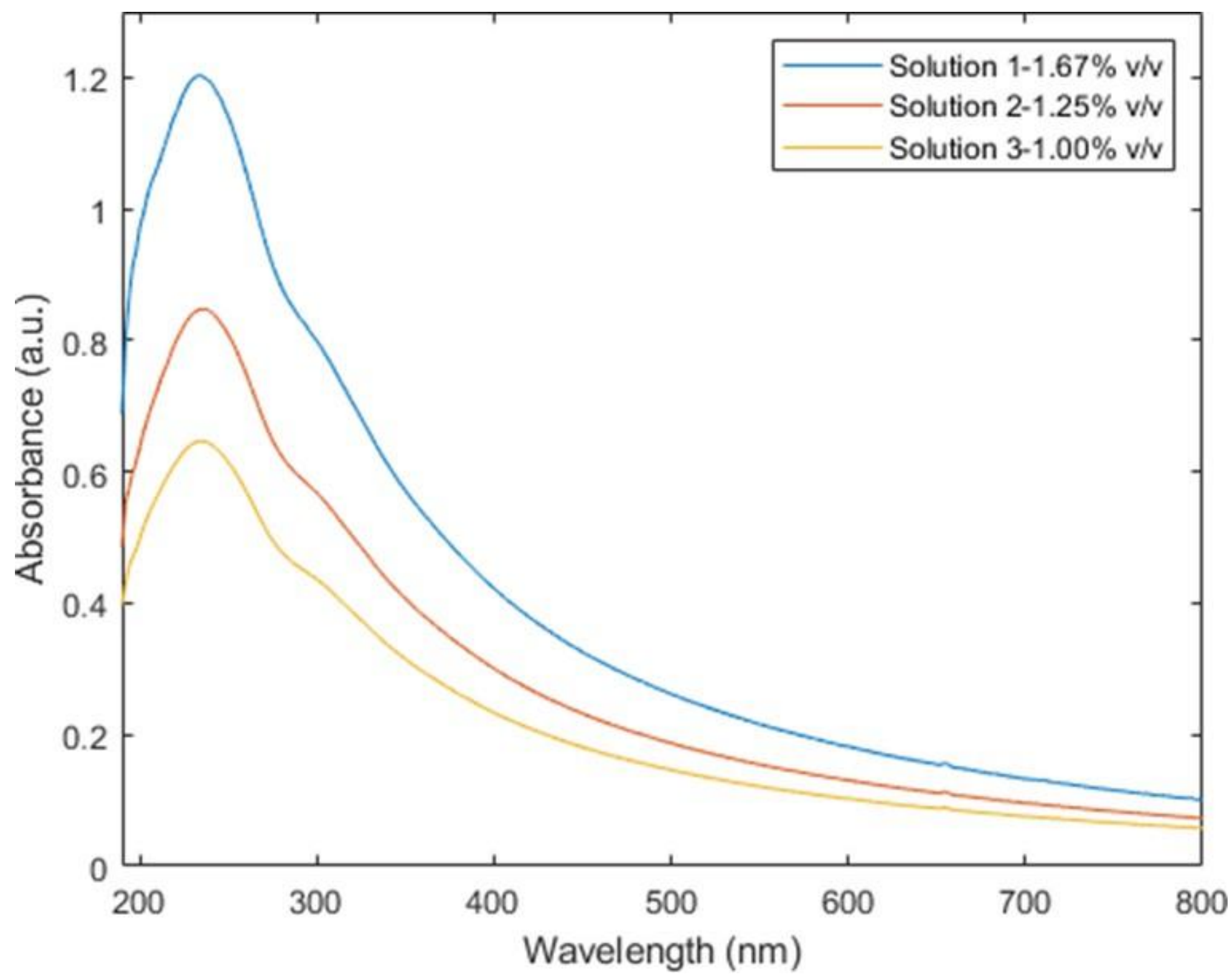

Figure 2: Absorbance of aqueous GO solution observed by UV-Vis spectrometer

It is evident that there is a relevance between the concentration of the solution and the absorbance peak; the higher the concentration of the higher the absorbance the higher the absorbance peak. This result is in accordance with the Beer-Lambert Law.

Table 1: Composition of aqueous GO solutions used for UV-Vis spectroscopy

| Solution | GO (mL) | Distilled water (mL) | GO content % v/v | Absorbance Peak (a.u.) | Wavelenth (nm) |
|---|---|---|---|---|---|
| 1 | 0.5 | 30 | 1.67% | 1.20 | 234 |
| 2 | 0.5 | 40 | 1.25% | 0.85 | 236 |
| 3 | 0.5 | 50 | 1.00% | 0.65 | 234 |

In all three cases the peak is observed on average at 235 nm, as well as the characteristic shoulder at 310 nm, both of which are aligning with literature. The slight deviation from the more standard 230 nm peak is expected, due to variation of the samples.

## 3.2 RAMAN SPECTROSCOPY

The Raman spectrum of the graphene aerogels that were fabricated using Freeze-Drying method was obtained at 532 nm in order to identify the material and to determine the level of oxidation of graphene oxide.

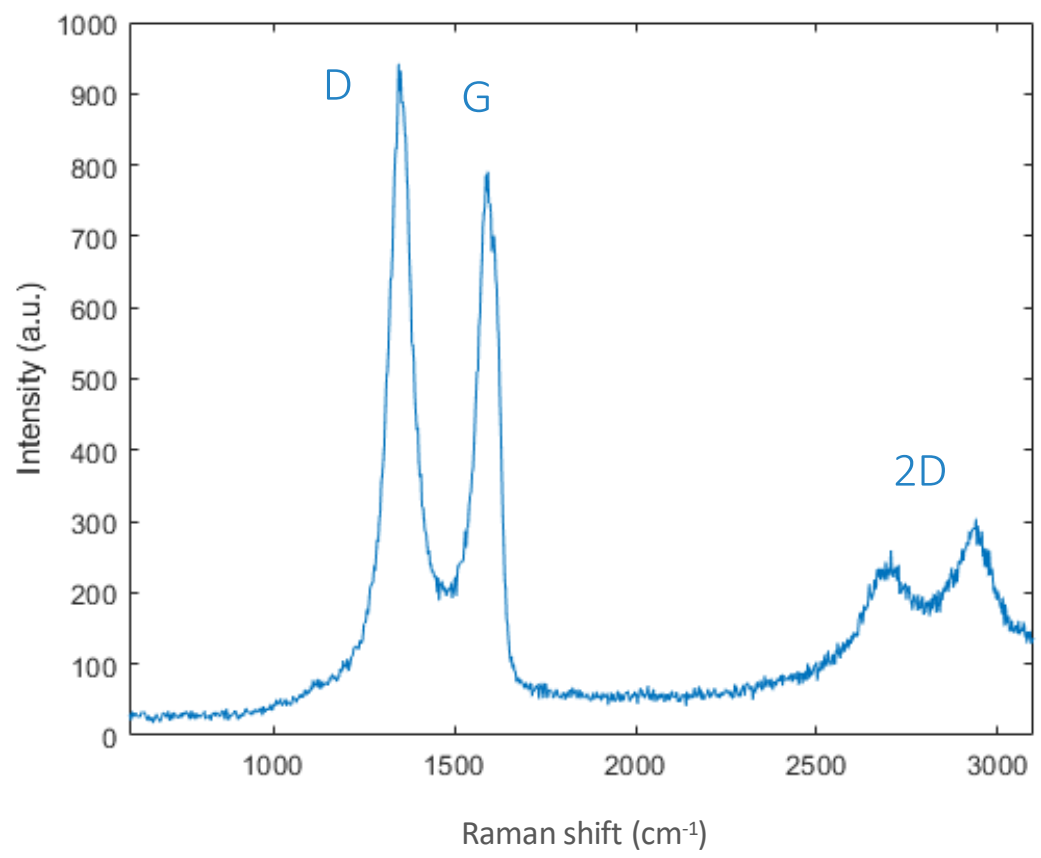

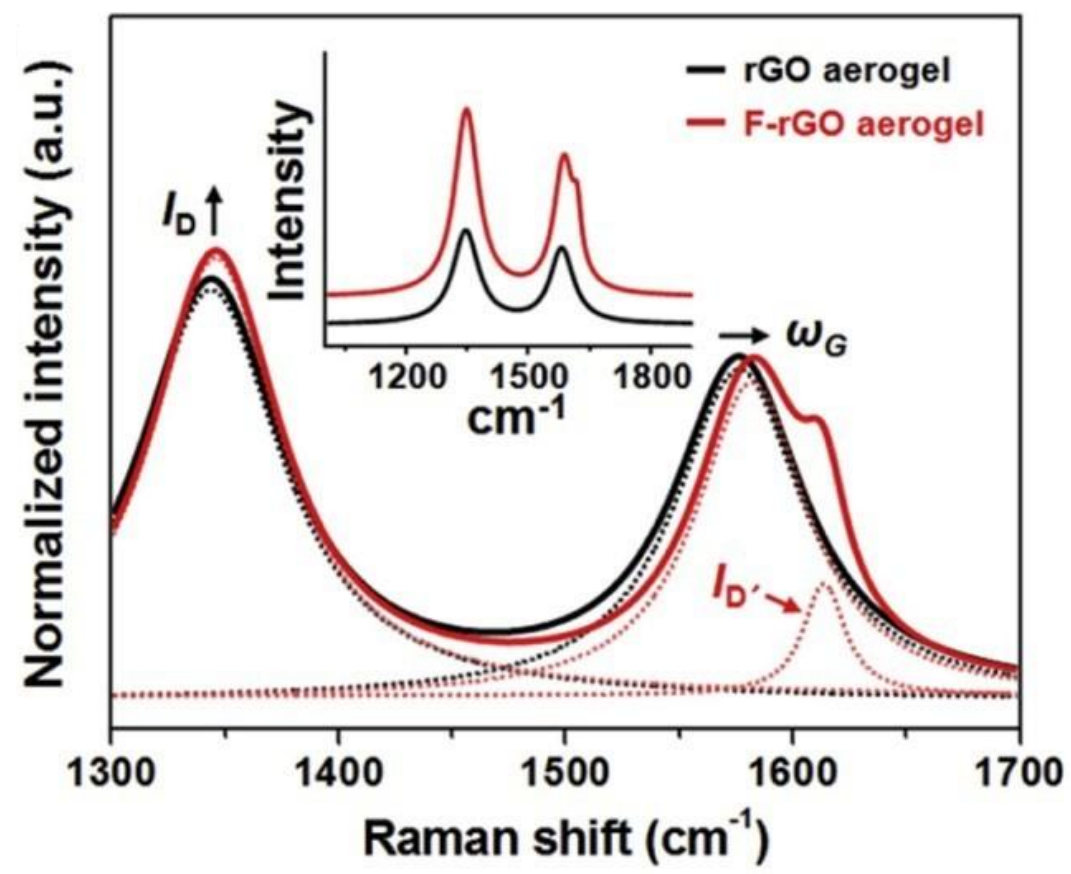

Figure 3: (a) Raman spectrum at 532 nm of the graphene aerogel prepared with FD method, (b) Raman spectrum obtained from Hong et al. [1]

The presence of D and G peaks of the sample are coherent to literature, which suggests that these are principal to graphene oxide and reduced graphene oxide. Theoretically, the D peak appears at approximately 1350 cm-1 and in this experimental case at 1351 cm-1. As mentioned before, it is the result of $sp^2$ bonds the adjacent to defects of the lattice and since the material examined is not monolayer graphene, it is expected that the peak won't be very defined. On the other hand, the G peak should appear at 1587 $cm^{-1}$ and in this spectrum, it appears at 1592 $cm^{-1}$. It originates from $sp^2$ hybridized C-C bonds in graphene materials. In the spectrum below, the 2D peak exhibits defined peaks at around 2812 cm-1. This shows that the reduction process of graphene oxide was successful and it demonstrates the restoration of the graphitic structure, as this band is always strong on graphene. The ID/IG ratio that quantifies the defects in the lattice and the conjugation disruption, was calculated at 1.17, while Hong et al. calculated it at 1.27. In comparison, the Raman spectrum of the graphene aerogels synthesized by APD method also exhibits the two characteristic peaks for graphene oxide and rGO, D and G. The G peak appears at 1641 and the D peak at 1841 cm-1 Raman shift. These peaks confirm the that the main, identifying bonds of the material are the sp2 hybridized bonds. However, contrary to Hong aerogels, here, the D peak has higher intensity, which may be attributed to higher defects of the lattice because of differ- ences in the fabrication process. Moreover, there is a distinct 2D peak, much like at FD GAs at, while the ID/IG ratio in this case is calculated at 0.85.

### 3.3 SEM

The SEM analysis of the samples gave information on their morphology and their pore sizes. The FD graphene aerogels exhibited pores in the order of magnitude of micrometer and some nanometers. It is evident from the obtained figures that layers of reduced graphene oxide with ripples and wrinkles are present. For reference, two pores of the material were measured in proportion to the μm scale bar of each SEM image. It is evident that the majority of the pores are micropores, thus confirming that the 3D material is microporous, similar to Hong's et al. [1].

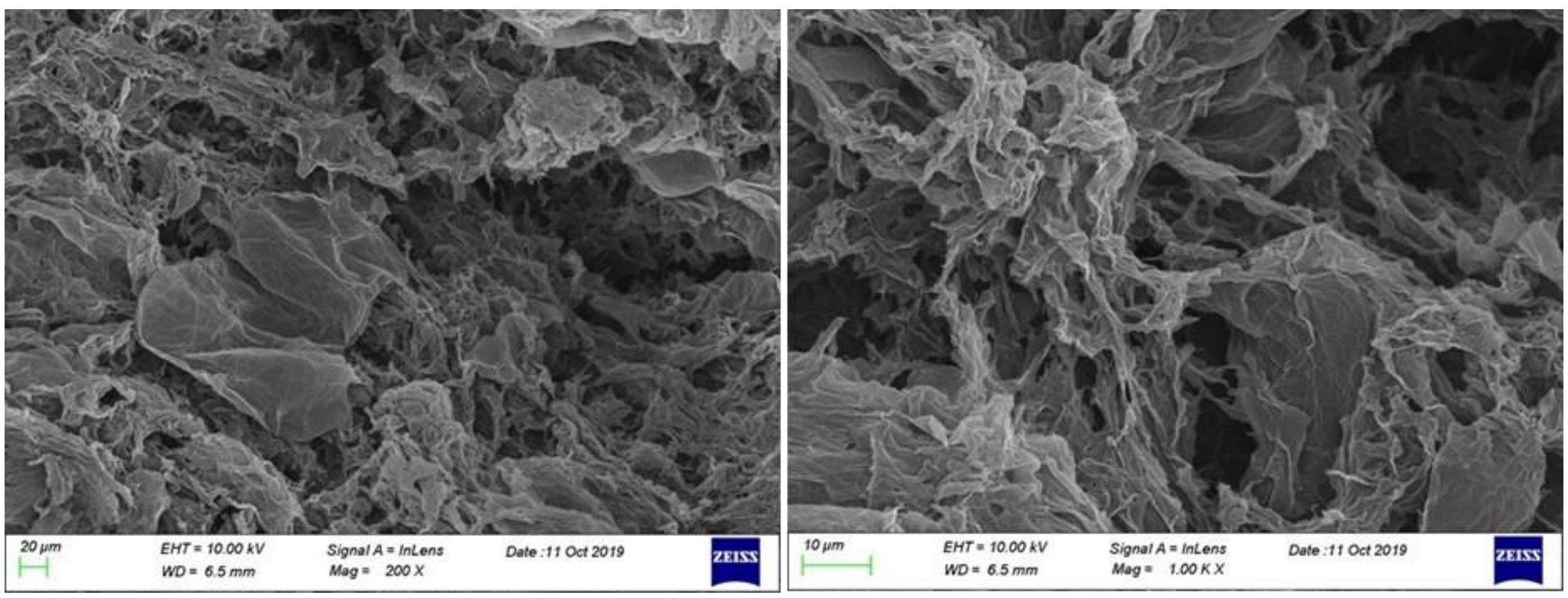

Figure 4: SEM images at 2 μm for graphene aerogels prepared with FD method

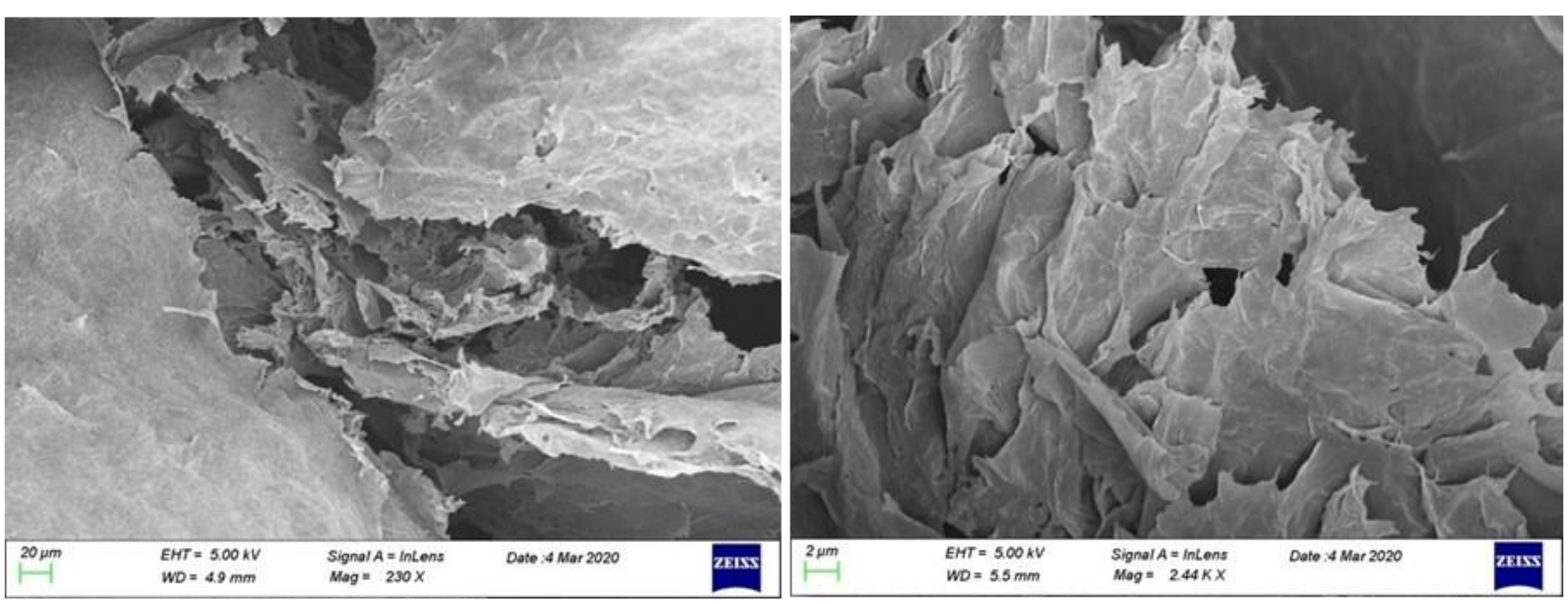

Figure 5: SEM images at 20 and 2 μm for graphene aerogels prepared with APD method

The survey suggests that most of the pores are micropores and the material can be characterized as microporous. This is coherent with literature [2], where all the prepared aerogels exhibit a macroporosity in their structure. Furthermore, the magnitude of the pores is noticeable even with the naked eye, thus, confirming visually the results. Some differences can be observed in the layers of reduced graphene oxide and the overall structure of the pores between APD and FD GAs, that may be attributed to the diverse chemical compounds and methods used in their fabrication.

### 3.4 VOCs absorption

The absorption capacity of the fabricated graphene aerogels was determined for the following VOCs: Formaldehyde, Acetic acid, Hydrochloric acid, Toluene. The absorption of the gas pollutants was measured initially in a gravimetric way. A quantity of 30- 40 mL of each selected volatile organic solvent was mounted in a desiccator. The graphene aerogels were put in a petri disks and their mass was measured on average every five days. The measurement was conducted with a high accuracy weight scale. The absorption capacity of the prepared graphene aerogels was determined by calculating the percentage of weight change comparing the lowest absorption to the highest one. Once the graphene aerogels reached the saturation point in the volatile gas, which corresponds to the maximum weight change, the weight of the GAs became noticeably lower. In order to establish a baseline for the GA saturation, after two consequent gravimetric measurements lower than the maximum observed value, the GA was considered saturated and was submitted to the regeneration process.

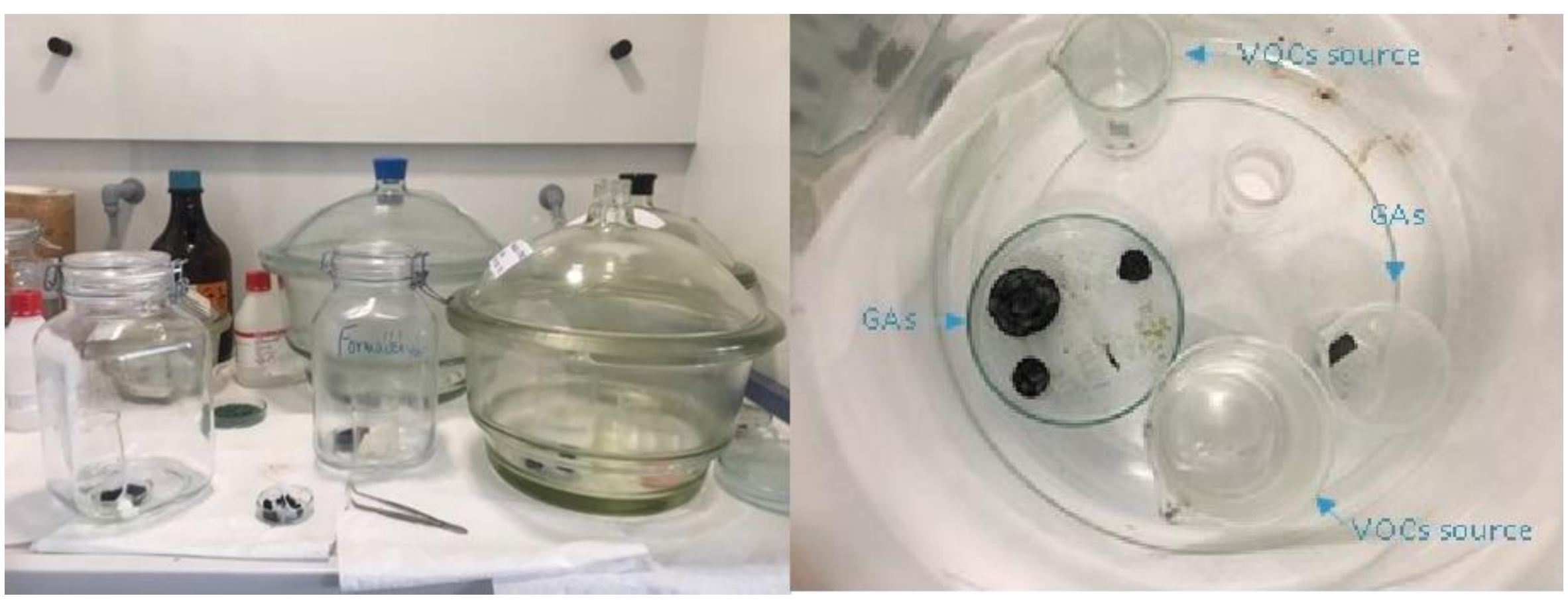

Figure 6: Experimental set-up for the measurement of VOCs absorption from the prepared graphene aerogels

The significance of the regeneration process lies in the necessity of recyclability of the 3D material and its capacity to be reused, from an environmental, economic and practical perspective. The re- generation was carried out according to existing literature that suggests the absorbent removal through drying above its boiling temperature. In the laboratory this was achieved with the use of a common electric hair dryer for 2-3 hours, as shown below:

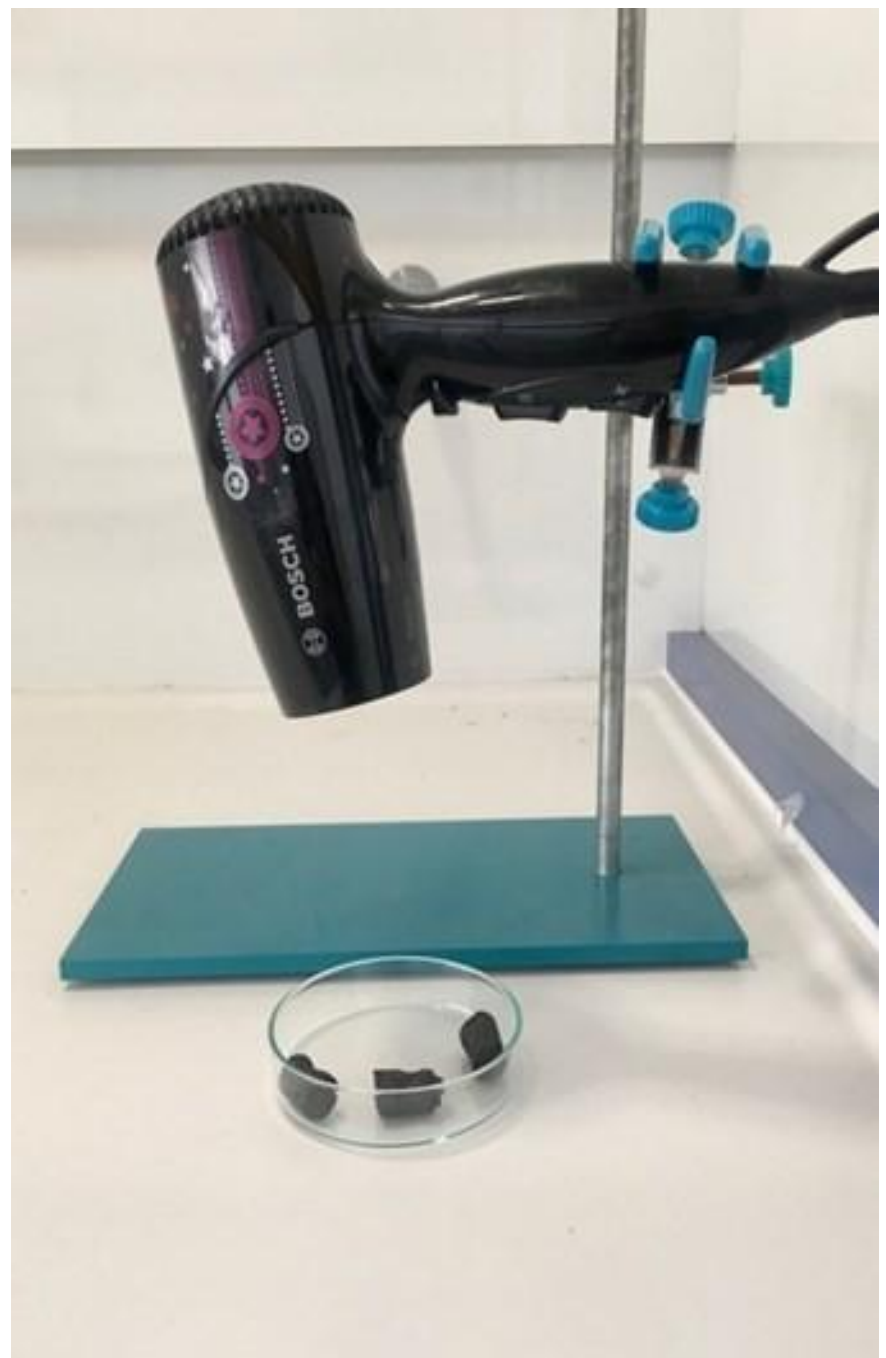

Figure 7: Experimental set-up for the regeneration of the graphene aerogels after reaching the saturation point

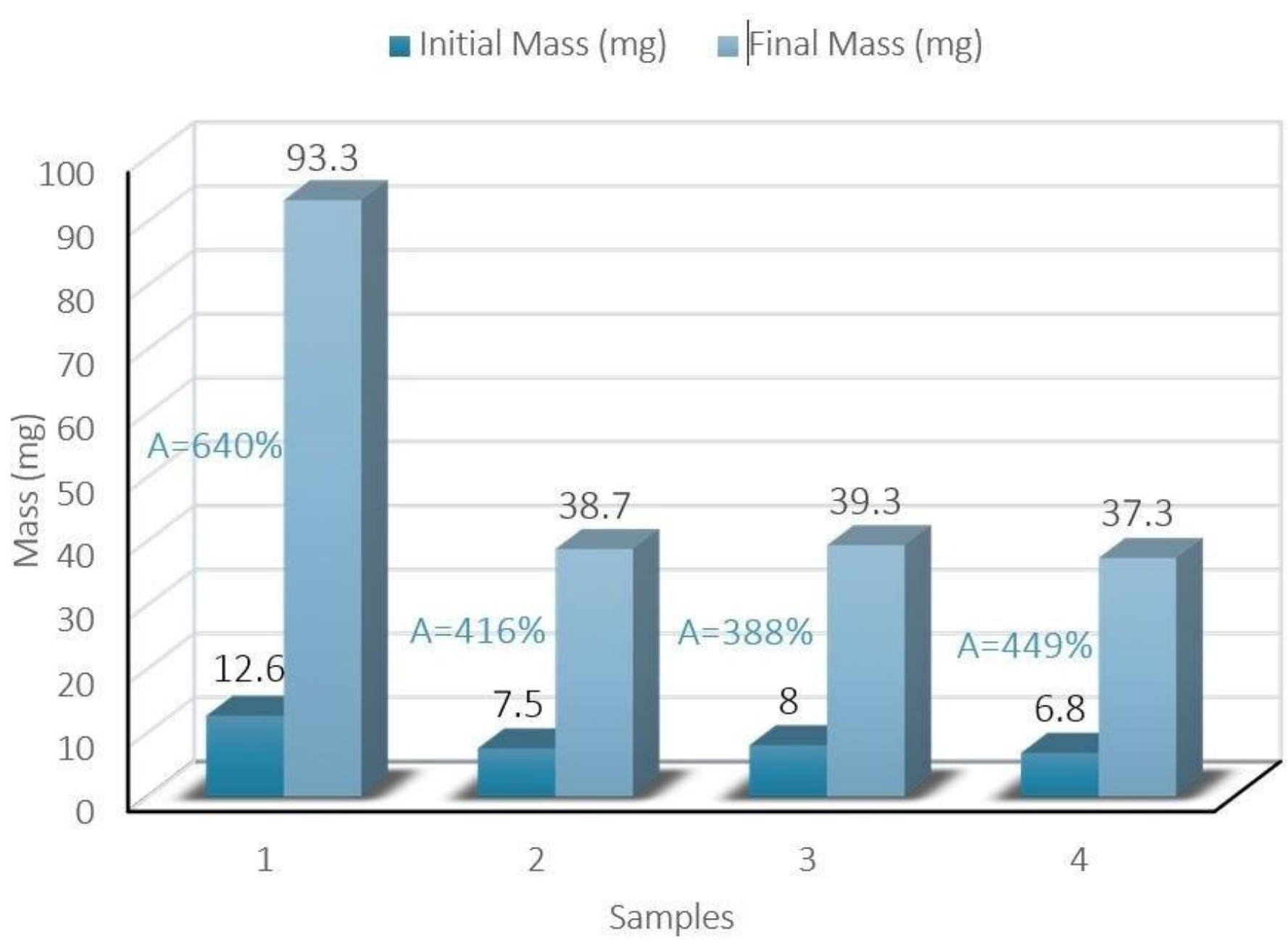

Figure 8: Mass difference of graphene aerogels prepared with FD method in a microenvironment of formaldehyde and A the % percentage w/w of VOC absorption during first round

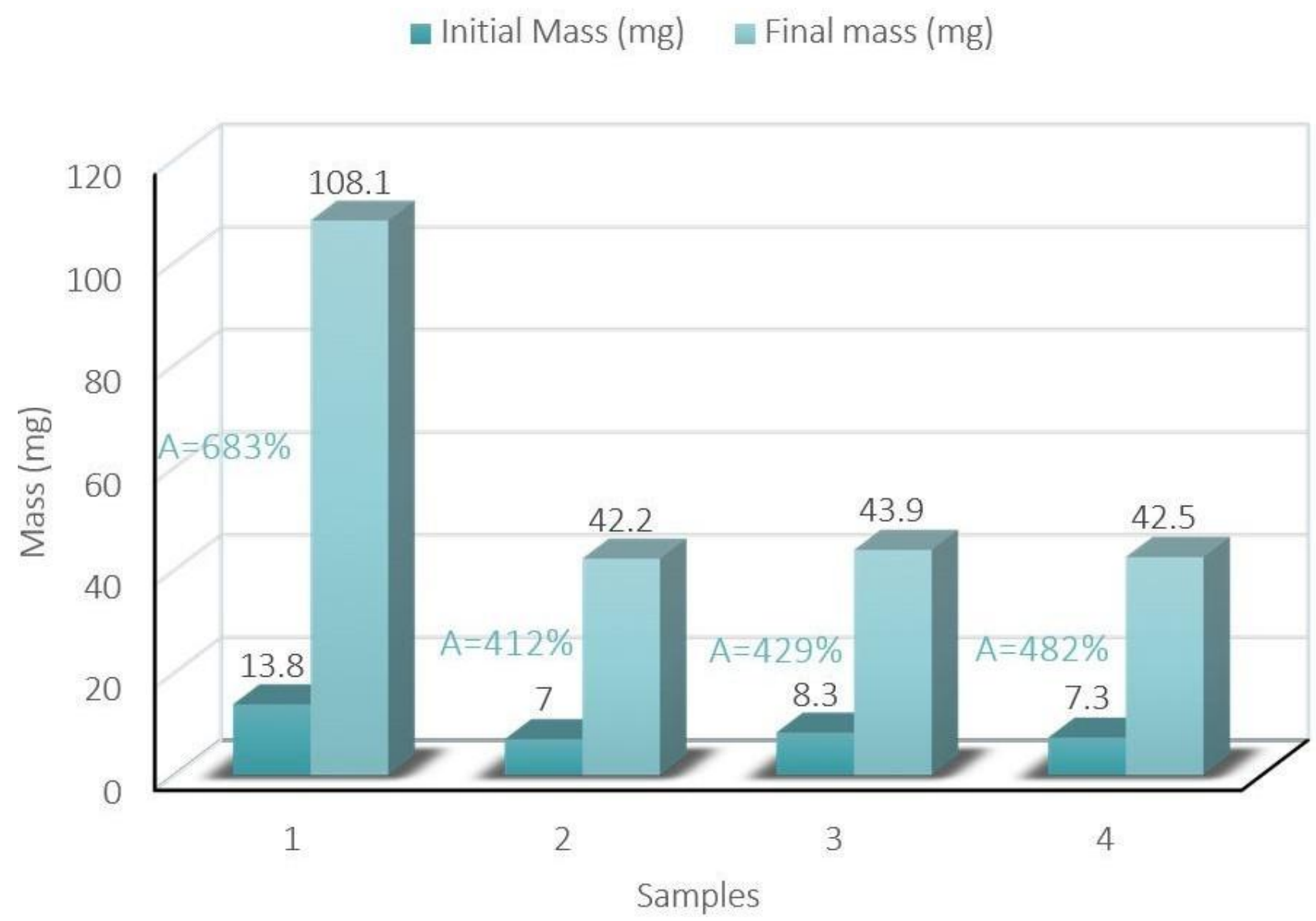

Figure 9: Mass difference of graphene aerogels prepared with FD method in a microenvironment of formaldehyde and A the percentage w/w of VOC absorption during second round

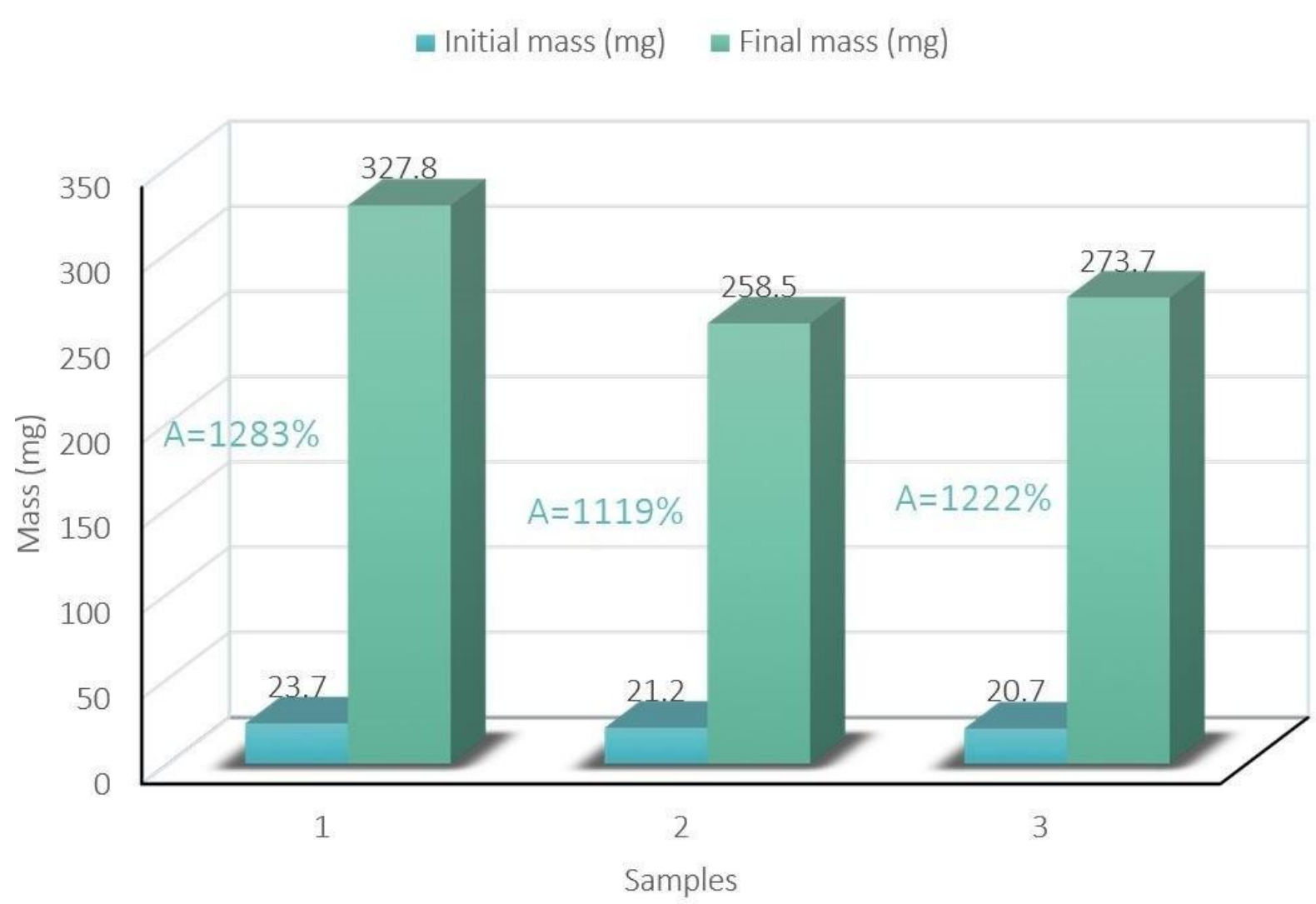

Figure 10: Mass difference of graphene aerogels prepared with FD method in a microenvironment of acetic acid and A the percentage w/w of VOC absorption during first round

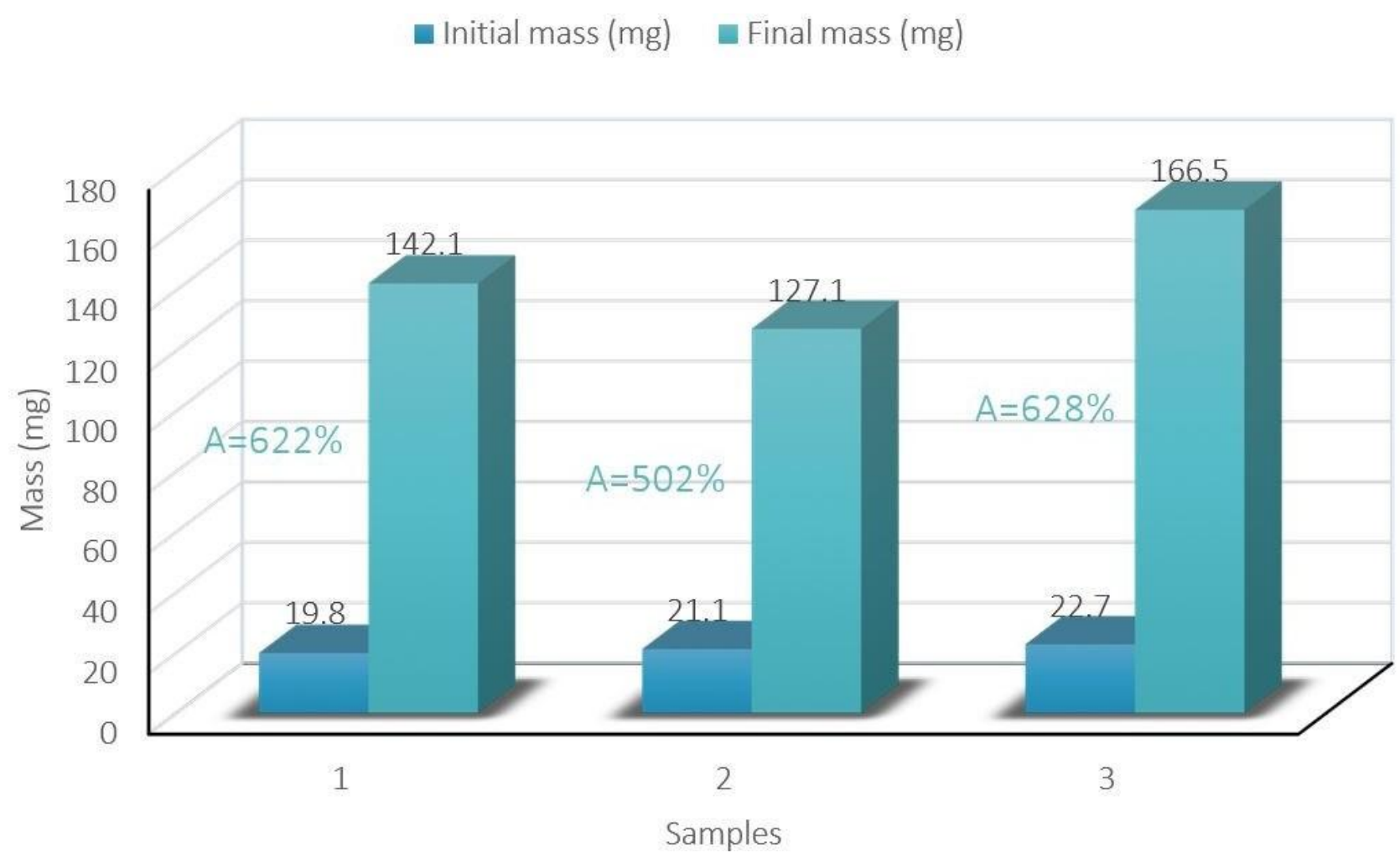

Figure 11: Mass difference of graphene aerogels prepared with FD method in a microenvironment of hydrochloric acid and A the percentage w/w of VOC absorption during first round

### 3.5 Measurements with VOCs sensor

Even though the gravimetric method of absorption measurement is commonly used and has proven efficient, it was decided to investigate the absorption capacity of the aerogel of the most significant VOC, formaldehyde, with a higher accuracy experimental set-up. Specifically, the experiment was assisted by a VOCs sensor. At first, the sensor was inserted in an almost air-tight vessel containing a petri dish of 30 mL formaldehyde. Once a plateau was reached, the microenvironment was considered saturated and two graphene aerogels of 2 mg/mL concentration were inserted as well. The aerogels were used in two rounds of absorption and the saturation point was defined as the plateau of measurements in a matter of hours. Each round had a duration of three days, in order to obtain data with the same time baseline.

During first round, the introduction of graphene aerogels in the formaldehyde microenvironment results in a noticeable decrease of voltage. This decrease of approximately 11.7% mV/mV supports the gravimetric measurements and the phenomenon of VOCs absorption from the GAs. The performance of the aerogels at the second round is significantly less efficient since there is almost no change in the plateaus with and without the presence of the GAs. This confirms the slower rate of absorption during second use that was observed in gravimetric methods as well. In the span of three days there is no significant absorption percentage of formaldehyde from the aerogels, therefore, a need is created for a longer sensor experiment in order to acquire the quality of results that offered the gravimetric method and to confirm the recyclability of the materials.

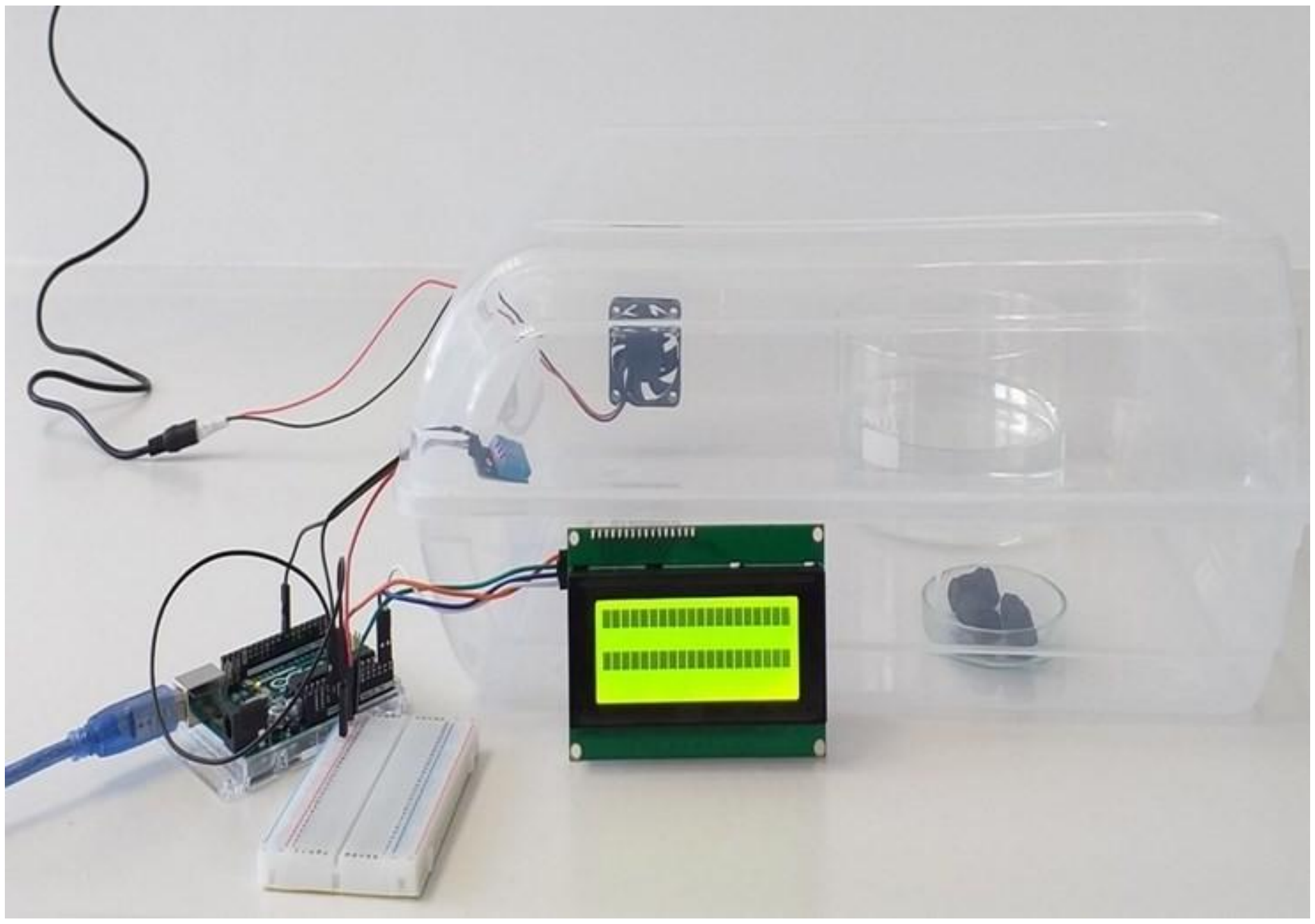

Figure 12: Experimental set-up with the VOCs sensor

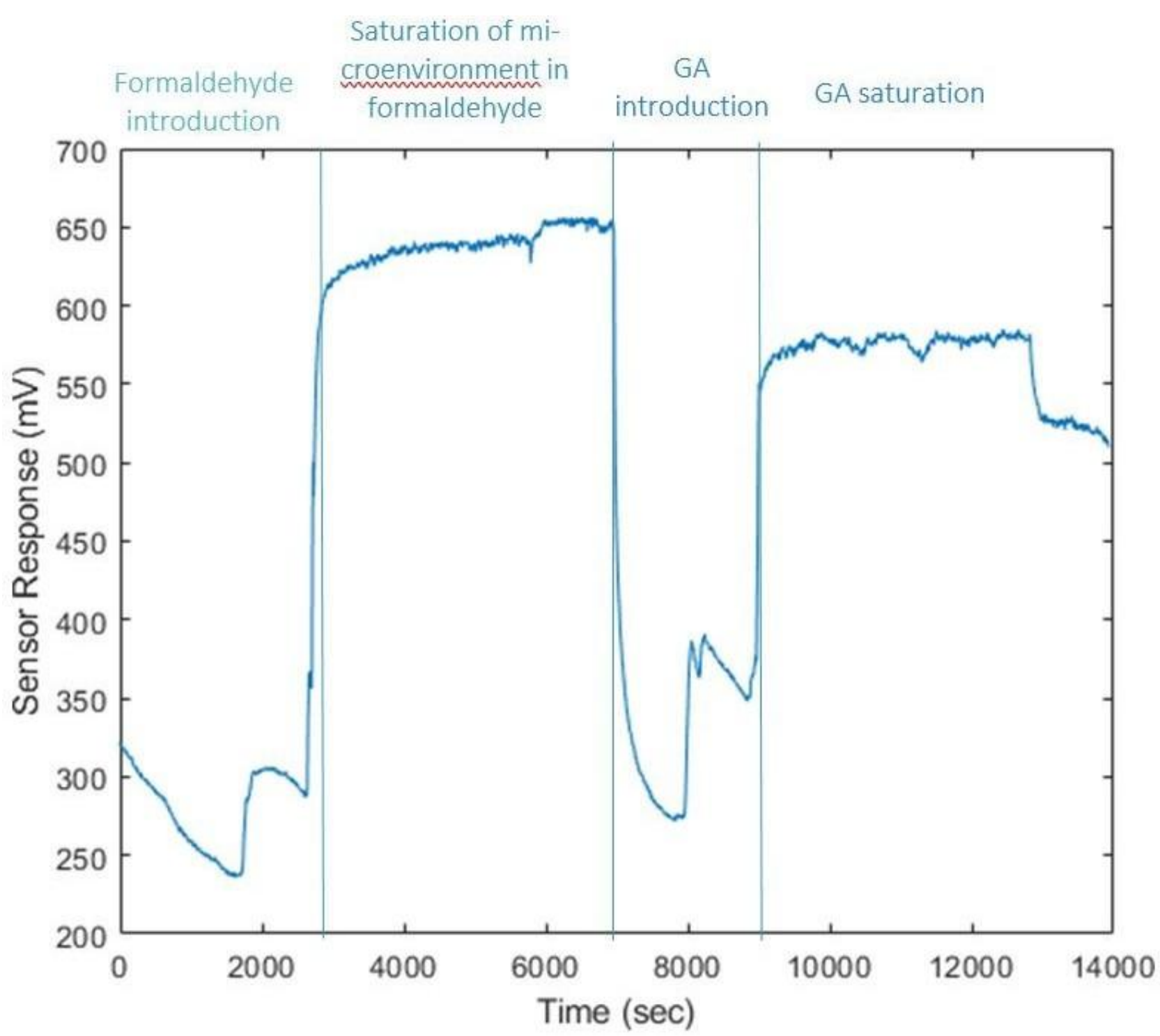

Figure 13: Graph of formaldehyde absorption from graphene aerogels prepared with FD method during first round

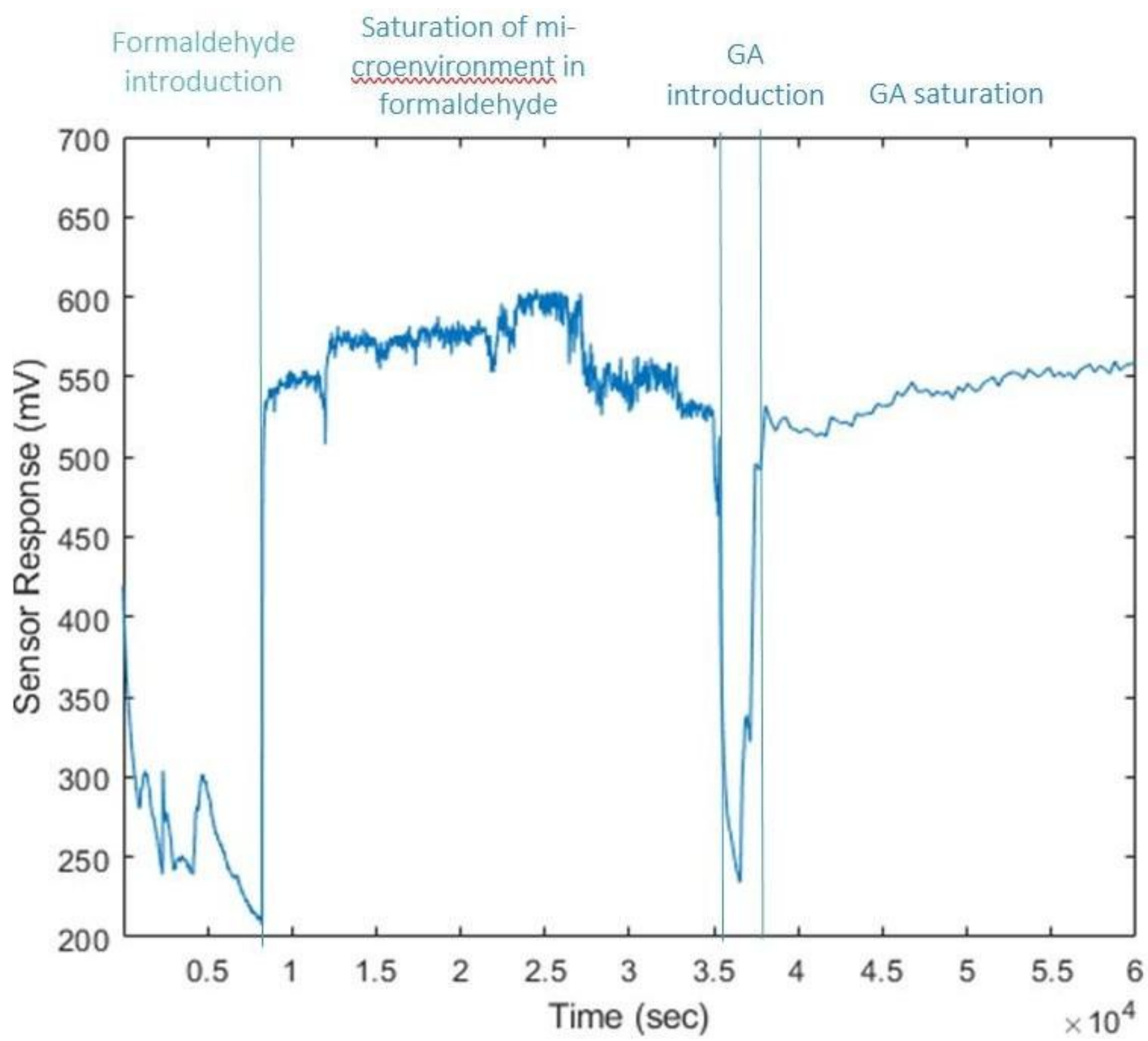

Figure 14: Graph of formaldehyde absorption from graphene aerogels prepared with FD method during second round

**3.6 Absorption of Organic Solvents**

In order to determine the absorption ability of the fabricated graphene aerogels regarding organic solvents, samples were soaked in ethanol, methanol and acetone and kept in an airtight vessel. The weight of the aerogels was measured with a high accuracy scale every two days until the saturation point. As saturation point was considered the maximum mass value and it was confirmed by two consequent measurements of smaller value. Graphene aerogels from both FD and APD method were tested to compare the absorption percentage w/w and notice any differences on the selectivity of the absorption. During the experiment, due to the volatility of the solvents the value of the gravimetric measurement constantly declined; therefore, the first and highest value was obtained as the most accurate.

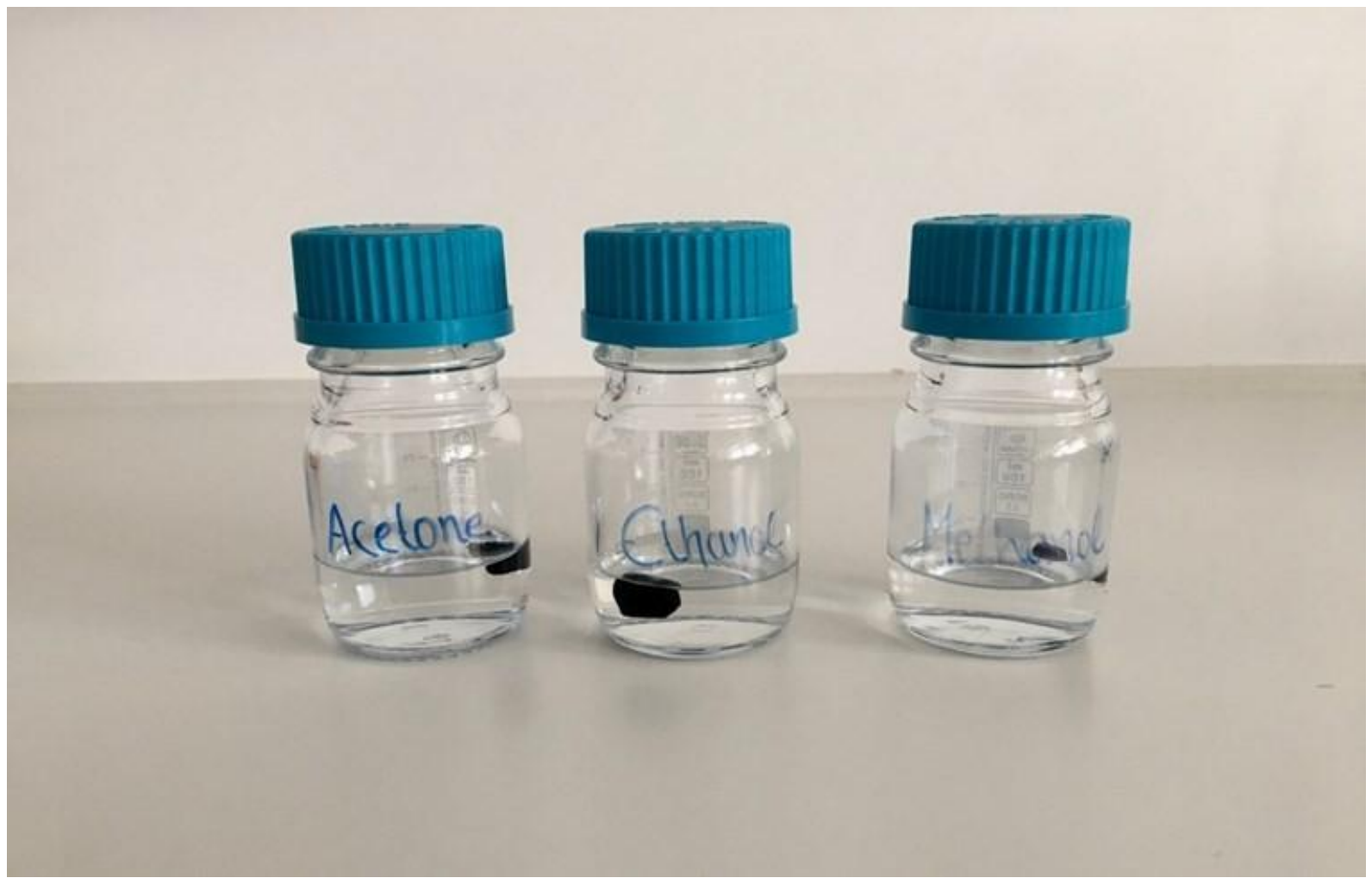

Figure 15: Absorption of organic solvents

The used aerogels reached the saturation point was reached in the time span of a week, while the overall the absorption capacity of the Hong aerogels is very high (over 5.000% w/w). From the graph below, it is deducted that even though the materials are very effective in absorbing the solvents, there is a slight selectivity toward methanol absorption.

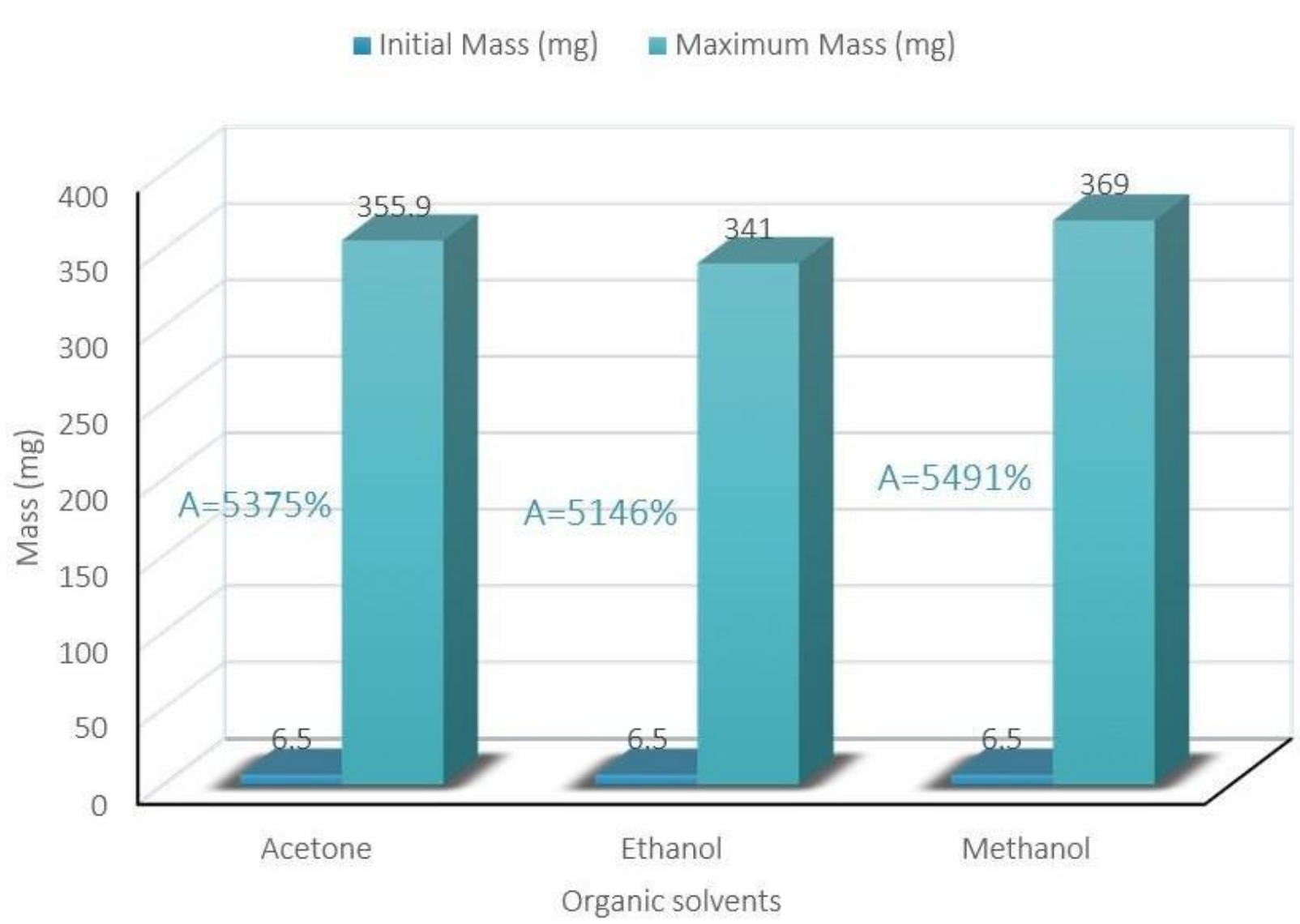

Figure 16: Absorption percentage w/w of acetone, ethanol and methanol from graphene aerogels prepared with FD method

## 4. Conclusions

The present study aimed at the synthesis of graphene aerogels with two different methods and the evaluation of the resulting materials as absorbents of liquid and gas pollutants. The synthesis of the aerogels was deemed successful, since the characterization techniques used showed that the reduction of graphene oxide was complete. The detailed characterization basically signified that the lattice of the material is almost identical to pristine graphene's; therefore, it exhibited the same absorbing properties but at a lower cost. The fundamental characteristic of both synthesis approaches, was that the reduction of graphene oxide was achieved chemically with the aid of heat. In FD method the main chemical agents were hypo-phosphoric acid and iodine, while in APD method the main chemical agent was L-ascorbic acid. The lack of thermal reduction lowered the energy consumption, as there was no need to use extremely high temperatures in the vacuum oven. The point of differentiation between the two methods was the use or lack of a freeze-dryer. The FD graphene aerogels were fabricated in the most common way as in by removing the water molecules from the hydrogel matrix by freeze-drying. On the other hand, Yang et al. employed a more facile approach of reducing the graphene oxide at steps, only using a freezer and the vacuum oven.

Comparing the efficiency of all the samples, it was evident that the graphene aerogels produced with FD method exhibited much higher absorption capacity, whereas the APD aerogels showed higher selectivity to VOCs and lower absorption percentages. The materials were examined in a gravimetric way as absorbents of pollutants and VOCs. The latter demonstrated significant absorption percentage and weight gain, which leads to the conclusion that the prepared aerogels could be used in application concerning the treatment of environmental emergencies.

**Funding:** This project has received funding from the European Regional Development Fund of the European Union and Greek national funds through the Operational Program Competitiveness, Entrepreneurship, and Innovation, under the call ERA-NETS 2019, SOLAR-ERA.NET (project title: Graphene cOmposites FOR advanced drinkingWATER treatment, project code:825207, T11EPA4-00090—GO FOR WATER, MIS 5070478).